\documentclass[conference]{IEEEtran}

\IEEEoverridecommandlockouts
\usepackage{cite}
\usepackage{amsmath,amssymb,amsfonts}
\usepackage{algorithmic}
\usepackage{graphicx}
\usepackage{textcomp}
\usepackage{hyperref}
\usepackage{subfig}
\usepackage{xcolor}
\hypersetup{
  colorlinks, linkcolor=red
}
\def\BibTeX{{\rm B\kern-.05em{\sc i\kern-.025em b}\kern-.08em
    T\kern-.1667em\lower.7ex\hbox{E}\kern-.125emX}}
\begin{document}

\title{A Systematic Literature Review on Internet of Vehicles Security\\
}

\author{\IEEEauthorblockN{ Priyankkumar Sharma }
\IEEEauthorblockA{\textit{Department of Engineering} \\
\textit{University of Guelph}\\
Guelph, Canada \\
priyankk@uoguelph.ca}
\and
\IEEEauthorblockN{ Meet Shitalkumar Patel }
\IEEEauthorblockA{\textit{Department of Engineering} \\
\textit{University of Guelph}\\
Guelph, Canada \\
mspatel@uoguelph.ca}
\and
\IEEEauthorblockN{ Apoorva Rajesh Prasad }
\IEEEauthorblockA{\textit{Department of Engineering} \\
\textit{University of Guelph}\\
Guelph, Canada \\
apoorvar@uoguelph.ca}
}
\maketitle

\begin{abstract}
The Internet of Vehicles (IoV), commonly referred to as connected automobiles, is a vast network that connects various entities, including users, sensors, and vehicles. They will connect across a network to lessen traffic accidents and improve both the security and safety of smart vehicles. The Internet of Vehicles is subject to a wide variety of threats, including spoofing attacks, recognition attacks, privacy attacks, and verification attacks. Our primary concern when creating any new smart gadget is the user's safety, which will be improved by identifying solutions to the various cyber threats. Therefore, we will cover the security of smart automobiles in this literature review, including their attacks and solutions.
\end{abstract}

\begin{IEEEkeywords}
VANET, IoV, V2V communication, security, 5G, cryptography.
\end{IEEEkeywords}

\section{Introduction}
Today’s fast-paced world demands an effective transportation system. Thousands of businesses rely on its transportation chain to cope with demanded supplies. A well-planned and secured transportation is a basic requirement of any organization. On the other side, transportation systems as well as personal vehicles are becoming costlier and inefficient to upgrade or maintain. A recent survey indicates that the worldwide number of vehicles is expected to be up to 2 million by 2035. Advanced technologies are being developed and upgraded to maintain new vehicles' efficiency and reliability \cite{b1,a1}. New paradigms such as cloud computing, embedded processors, IOT based models are recent advancements in the development of intelligent devices. The increased number of inter-connected vehicles conceived the concept of a vehicular ad-hoc network (VANET) and internet of vehicles (IOV).
VANET is a basic model to connect and stimulate vehicles in a specific range, it can not evaluate and update global real-time data. On the other hand, the internet of vehicles is equipped with networking and intelligence.\cite{b2} IOV integrates other objects such as humans, vehicles, networks, traffic, and routes to develop efficient and reliable services. Connection of vehicles and other cyber-physical components such as sensors, internet, and satellites, can provide a global network equipped to evaluate all real-time data which leads to a safer, more efficient, and reliable world of transportation\cite{b4,a2}
\begin{figure}[htbp]
\centering
\includegraphics[width=10cm,height=8cm]{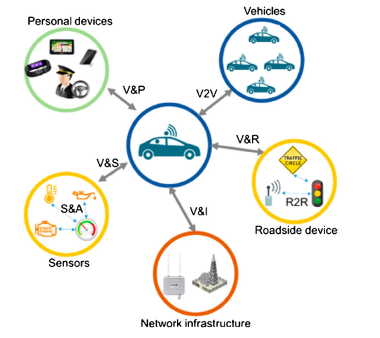}
\mbox{}
\caption{Communication in IOV\cite{b2}}
\end{figure}
\begin{figure}[htbp]
\centering
\includegraphics[width=8cm,height=8cm]{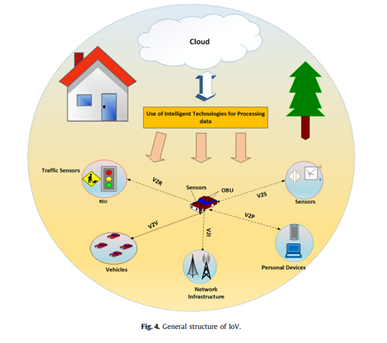}
\mbox{}
\caption{General Structure of IOV \cite{b2}}
\end{figure}
To continue, our major focus in this SLR is to identify possible threats and possible enhancement of security measures in IOV. A basic secured IOV can be developed by ensuring security aspects such as authentication, data integrity, confidentiality, end-to-end encryption, and access control \cite{a3,a6}.
IOV is highly vulnerable to attackers, it can be hacked and manipulated by simply changing its vehicle to a server connection which can cause tragic accidents, data breaches, and loss of highly valued information.\cite{b6,a3} There are multiple other security measures related to each layer of IOV which we will discuss further in our SLR. Different types of attacks are categorized according to its characteristic and relative components in IOV which we will further discuss.
\subsection{Basic research}
The Internet of Vehicles is a vast field that advances day by day to continuously improve and fulfill challenging transportation needs. Our focus in this systematic literature review is on how we can ensure security measures and find a feasible solution against possible attacks.\cite{b5} There are multiple articles and research papers available on architecture, connectivity, users, and stakeholders in IOV. There are limited descriptions, methodologies, and solutions available for possible threats and attacks in IOV. \cite{b4}
In 2018 Yosra Fnon-Englisha detailed a description of present issues in the security of IOV. He provided information about cyber attacks and how different components of IOV can be attacked and which attack will compromise the security of specific components.\cite{b1}
\begin{figure}[htbp]
\centering
\includegraphics[width=8cm,height=5cm]{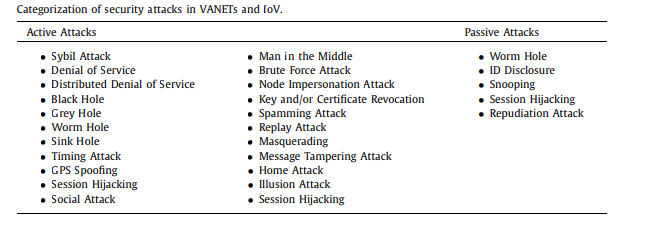}
\mbox{}
\caption{Types of Cyber Attacks}
\end{figure}
Surbhi S. published an article in 2019 for Vehicular Communications. By identifying its applications, architecture, and security issues mentioned we can say that it is the most informative article related to IOV. Different possible attacks are mentioned in the table, and we aim to find an effective solution for each of them which can motivate ongoing studies to advance the security of the Internet of vehicles.\cite{b2}
\subsection{Purpose of research}
This research aims to identify and analyze ongoing studies and existing literature on the security of the Internet of vehicles \cite{b2,a7}. Moreover, our purpose also includes measuring the feasibility and effectiveness of existing solutions for cyber security for IOV.

\subsection{Contribution and Layout}
Up to the beginning of 2020, we identify 24 primary research on IoV and IoT security. This list of studies can be used by other scholars to advance their own work in this particular area. 13 main studies that satisfy the standards we set for quality evaluation are further chosen. These papers can serve as appropriate comparison points for the examination of related research. In order to represent the research, concepts, and considerations in the disciplines of blockchain and cyber security, we undertake an in-depth analysis of the data included in the subset of 13 papers

The format of this essay is as follows: The techniques used to choose the primary studies for analysis in a methodical manner are described in Section 2. The results of all the primary research chosen are presented in Section 3. The findings in relation to the prior study topics are discussed in Section 4. Section 5 wraps up the study and makes some recommendations for more research.
% \begin{figure}[htbp]
% \centering
% \includegraphics[width=8cm,height=4cm]{question table.jpg}
% \mbox{}
% \caption{Cyber Attacks}
% \end{figure}
\section{Methodology}
We carried out an SLR in this survey, which entails planning, conducting research, and reporting. The research questions were established during this planning phase, and the rules and approach will be established during the search phase. The outcomes were finally presented \cite{b3,a10}. The purpose of this survey is to identify the dangers and vulnerabilities that exist within IoVs.
The following issues for more study were found:
RQ1: What are the different IoV risks, and whose security service are they affecting? This question is meant to highlight the need of being able to categorize each threat to effectively address it.\cite{b7}
RQ2: What are the numerous options that may be used to deal with the dangers mentioned in RQ1? To choose the response that best meets our needs, we must consider every alternative to the question.
RQ3: What effects does each solution have on the operation of the system? The objective is to guarantee that, when utilizing any particular solution, performance won't degrade below what is anticipated.
\subsection{Selection of Articles and Papers}
The advanced search phrase ("IoV" OR "Internet of vehicle security" OR "VANET" OR "Intelligent transportation systems" OR "In-car Internet" OR "Connected automobiles" OR "V2V communication") was used to find information for the aim of addressing the stated research objectives ("Threats" OR "Vulnerability "OR "Solutions"). The inclusion of any internet sources that deal with IoV, Internet of Vehicle Security, VANET, Intelligent Transportation Systems, In-Car Internet, Connected Cars, or V2V communication will be followed by filtering based on the sources' dangers, vulnerabilities, or remedies. The aforementioned digital libraries were searched to locate the required publications (from journals and research papers):\cite{b4} \\
•	IEEE Explorer\\
•	Google Scholar\\
•	Science Direct\\
•	ACM Digital Library\\
•	Springer\\
\subsection{Inclusion and exclusion Criteria}
As a result, when discussing the inclusion and exclusion criteria, we shall discuss the types of papers that must be included and those that must be excluded. Therefore, the papers we will submit for this SLR must address topics such as IOV Internet of Vehicle security, VANETS applications, Smart Cars, and anything else involving automobiles.\cite{b13}

Criteria for Inclusion\\
-	Paper must be related to the security of that car.\\
-	Paper must include applications of the VANETS, IOV.\\
-	Only case study papers are valid.\\
Criteria For Exclusion\\
	 - Remove the non-English paper.\\
	 - Not having valid findings.\\
	 - Not related to the topic. \\    
\subsection{Selection results}
We go over our findings on the earlier defined study questions in this part. In a survey bathatarticles from 2010 to 2022, the IoV's (or VANET) protocol covers security risks, solutions, and how the performance will be affected by a solution.\cite{b4}
To prevent endangering the lives of individuals who use them, we must ensure that new technologies are as safe as they possibly can be in a connected world. The majority of papers describe the VANET protocol, which is used for communication.
IoVs rely on wireless technology to operate, making them susceptible to various threats like jammers. For instance, jamming creates interruption by producing a signal that is identical to the signals produced by the vehicles \cite{b10}.
\subsection{Different threats in IoV}	 
The IoV's broadcast nature makes cars simple to target, and the constant movement makes it more challenging to follow the attacker, therefore we need a secure protocol and a system that enables the cars to connect safely and covertly.
Like any other system, we aim to include the following four key security components in ours:
1. Integrity: Assuring that the communicated data is accurate, error-free, and has not been modified during transmission by a hostile attacker; using hashing techniques is an easy method to do this.
2. Authenticity: Ensuring that the message's sender is who they say they are and is not another person posing as them.
3. A simple example of this is encryption. Confidentiality: This is similar to privacy in that we must make sure that sensitive data is safeguarded and that only authorized users may access that data.
4. Accessibility: ensuring thatnon-English is constantly accessible, faultless, and functioning as it should \cite{b4,j1}.
\begin{figure}[htbp]
\centering
\includegraphics[width=8
cm,height=6cm]{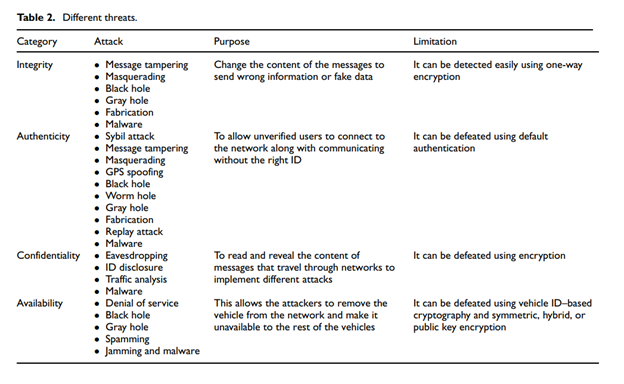}
\mbox{}
\caption{Cyber Attacks}
\end{figure}
\subsection{Measurement of Quality and Data Scrubbing}
After a thorough examination of the research papers, it could be noticed that the attacks on IoV have different types. Some of the major attacks could be identified as authentication attacks, secrecy attacks, routing attacks, and data authenticity attacks. There are more threats in IoV but this keyword is repeated most commonly. Apart from this other threats that are more generic and are applicable in most cases are denial\cite{11} of services and impersonation attacks, which could be used to get improper authorization and access. All these threats were taken into consideration and are given more priority for the research purposes of this paper.
\section{Findings}
Every research paper was analyzed and summarised according to the key application and purpose of the research. As mentioned in the research methodology we started the search with keywords context of security on the Internet of Vehicles and found different applications and their security challenges. Table NO shows the findings of papers with key qualitative and quantitative data and their key applications \cite{b10,j2}.

Research shows different possible attacks in IoV and its possible solution. There is a vast area of application of IoV(e.g. AI-based, Blockchain, Communication, federated learning) which also require a protected and secure environment for data transaction. In this research, we discussed the common attack which can compromise the security of the whole network layer and data transportation. Figure NO shows the findings of attacks discussed in various research papers.

\begin{table}[htbp]
\caption{The key research' main results and topics}
\label{Table3}
\setlength{\tabcolsep}{10pt} % Default value: 6pt
\renewcommand{\arraystretch}{2.5} % Default value: 1
\begin{tabular}{|p{1cm}| p{14em} | p{1.5cm}|}
\hline
\textbf{Primary Study} & \textbf{Key Qualitative \& Quantitative Data Reported}& \textbf{Types of Security Measurement}\\ 
\hline
\cite{b14} & The growth of the internet of vehicles is accelerating as it links the occupants of the car to various cyber systems that can be used to scan for and identify viruses. By connecting vehicles, sensors, and mobile devices into a worldwide network, IoV extends beyond telematics, vehicle ad hoc networks, and intelligent transportation to enable the delivery of numerous services to vehicular and transportation systems, as well as to persons inside and around vehicles. & IOT Security Attack   \\
\hline
\cite{b15} & The processor will essentially function as a little computer on board, with adequate processing capability to handle both the communication modules installed on board and the processing of incoming data & IOV Security Attack\\
\hline
\cite{b16} & Intravehicular communication protocols in VANETs are critical in IoV because they enable various levels of interaction between vehicles, humans, and roadside equipment. If there is a difficulty with the current route, they can suggest alternate routes efficiently and swiftly. & 
IOV Security Attack
\\
\hline
\cite{b17} & CAV cybersecurity can be handled logically on multiple levels: Vehicle, Infrastructure, and Human Factors & 
Spoofing, DoS
. 
\\
\hline
\cite{b18} & By enabling cooperative communication between autos, roadside infrastructure, and traffic management centers, VANET can increase traffic efficiency and safety. & Man In the Middle. 
\\
\hline
\cite{b19} & Developing telecommunications promotes technology transformation it helps numerous industries, including education, energy, smart cities, and connected cars. Due to the enormous development potential of smart transportation, IoV is attracting increased interest from a growing number of academics.  & AI Security. 
\\
\hline
\cite{b20} & The entity with the greater reputation value is only permitted to pack the resource transactions in this method, and the reputation values are maintained on the blockchain. The reputation values for each entity are determined by taking into account both current computational performance and historical history. In this context, a game theory-based system is devised to deliver incentives to participants. & BLOCKCHAIN Security. 
\\
\hline
\end{tabular}
\end{table}

\begin{table}[htbp]
\label{Table3}
\setlength{\tabcolsep}{10pt} % Default value: 6pt
\renewcommand{\arraystretch}{2.5} % Default value: 1
\begin{tabular}{|p{1cm}| p{14em} | p{1.5cm}|}

\hline
\cite{b21} & V2H (Vehicle-to-Human) or Vehicle-to-Pedestrian (V2P) communication links both vehicle and humans (passengers, drivers, etc.) via smart devices (smartphone, tablet, smart watch, etc.) via Apple's CarPlay or Android's OAA system or Near Field Communication (NFC) [23], [24]. It is based mostly on cellular technology (3G, 4G, and 5G). The IoEV enables electric vehicles to communicate with the Internet. This communication might take several forms (V2V, V2I, V2H, V2N, V2S, EV2EVSE, and EVSE2N). & Black Hole Attack   \\
\hline
\cite{b22} & Third-party systems (TPS) handle all connection management and priorities wireless technologies for connection setup. TPS is a service level commitment that is difficult to implement due to the various and heterogeneous nature of IoV networks. & IOV Security Attack\\
\hline
\cite{b23} & The cloud system either creates an autonomous cloud of cars or links the vehicles to a standard cloud. In any instance, the resources of connected vehicles may be used as a cloud service, and the vehicles can use smart cloud services. The solution will reduce automobile storage and computation constraints. It might pave the way for new commercial models in linked driving. & CLOUD COMPUTING
\\
\hline
\cite{b24} & Wireless nodes in vehicle-mounted networks behave differently. Some of these characteristics may have an impact on the design of IoV technology. & 
ITS
\\
\hline
\cite{b25} & By enabling cooperative communication between autos, roadside infrastructure, and traffic management centers, VANET can increase traffic efficiency and safety. & Man In the Middle. 
\\
\hline
\cite{b26} & In an IoV network, all vehicles are obliged to maintain a record of the trust ratings of every other vehicle they have interacted with during their individual trajectories.  & IOV Security Attack. 
\\
\hline
\cite{b27} & As opposed to feed-forward neural networks, LSTM is a kind of artificial recurrent neural network (RNN) deep learning model. It has uses in time-series data fields including speech recognition, handwriting recognition, and network traffic intrusion detection systems. & Sybil Attack. 
\\
\hline
\end{tabular}
\end{table}

\begin{table}[htbp]
\label{Table3}
\setlength{\tabcolsep}{10pt} % Default value: 6pt
\renewcommand{\arraystretch}{2.5} % Default value: 1
\begin{tabular}{|p{1cm}| p{14em} | p{1.5cm}|}

\hline
\cite{b28} & In general, the phrase "machine learning" refers to computerized methods that essentially collect data and data streams and seek for patterns to fast respond depending on what the algorithms are learning over time. & DoS Attack   \\
\hline
\cite{b29} & Datasets and codes produced and/or analyzed during the current investigation are accessible upon reasonable request from the corresponding author. & DEEP LEARNING\\
\hline
\cite{b30} & In Wireless sensor networks, machine learning (ML) enables intelligent forecasting and decision-making by assisting with huge data analysis.      & ML
\\
\hline
\cite{b31} & Value-added data may violate someone's privacy since it is gathered, examined, and kept by several organizations in the SIoV architecture. This study examines each layer of the SIoV architecture in order to go deeper and uncover the fundamental causes of various privacy issues. & DoS Attack
\\
\hline
\cite{b32} & To establish if the car identity is genuine or not, the cloud service providers mostly employ the Rayleigh consensus process and query the presence of the vehicle identity ID. & IOV Attack. 
\\
\hline
\cite{b33} & The vehicle is able to determine its current position, speed, and direction in degrees thanks to GNSS and Location Service Unit.  & GNSS. 
\\
\hline
\cite{b34} & A paradigm for threat modelling and risk assessment for autonomous cars is provided by SARA (Security Automotive Risk Analysis). Despite being a tool for assessing security risks, it allows for the examination of safety problems brought on by security concerns. & SARA Attack. 
\\
\hline
\cite{b35} & Through WSN, data and information are gathered and communicated, and attackers actively and aggressively target data or objects with a WSN connection. & IOT Attack. 
\\
\hline
\cite{b36} & Data immutability, data timeliness, and data security problems can be solved with the use of distributed ledger technologies. & IOTA Attack. 
\\
\hline
\cite{b37} & Failure mode effect analysis (FMEA) based on the lessons gained is one of the additional phases in recovering from such assaults. & FMEA. 
\\
\hline
\cite{b38} & The techniques are grouped in the papers acquired from the research in terms of security, reputation, secrecy, decentralization, trust-based approaches, data sharing, and authentication, among other factors. & BLOCKCHAIN Attack. 
\\
\hline
\end{tabular}
\end{table}

\begin{figure}[htbp]
\centering
\includegraphics[width=9
cm,height=6cm]{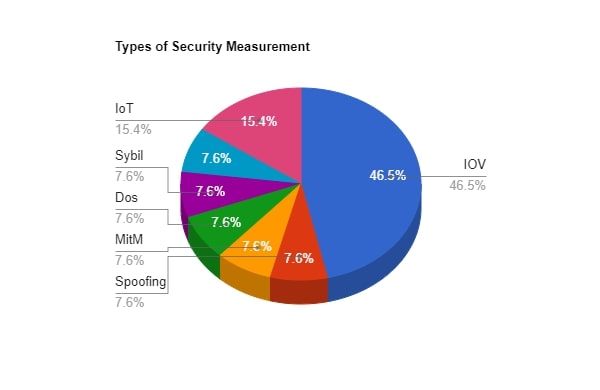}
\mbox{}
\caption{Pie chart representation of cyber attacks}
\end{figure}

According to the pie chart, the IOV attack accounts for the greatest number of papers (46\%), followed by IoT (15.4\%), and the remaining articles (7.6\%) are about Sybil, DoS, MITM, and Spoofing. Consequently, IoT and IOV security was primarily the emphasis. If we look at the table, we can see that I pulled 24 articles from it, all of which are based on IOV and IoT. When I checked Google Scholar, I found that most papers on the IoT are about security and attacks, which is what I am searching for. Other than this one, the remaining articles are founded on Sybil, MitM, Dos, and spoofing.

\section{Discussion}
Due to the IoV's wide-ranging potential, an increasing number of nations and organizations are studying how to integrate intelligent transportation with traditional transportation. This has led to an increase in unanticipated security issues, which also draws attention to the issue. Since 1999, the United States has published Fair Information and Privacy Principles for its Intelligent Traffic System (ITS), and the National Institute of Standards and Technology created the Cybersecurity Risk Management Framework Applied to Modern Vehicles [1]. Additionally, the European Union launched its ITS Action Plan to limit the use of IoV data and safeguard its security. So, the need for security features in IoV protocols will rise as the technology develops and becomes widely used shortly. To ensure that the IoV platform adheres to the regulatory requirements for the security and privacy of users and vehicles, distributed, scalable, and strong security solutions are needed. Although weIn VANETs and IoV, the Sybil assault is one of the most common types of attacks. In conducting our research on the Sybil assault, we consulted recent reference publications. To prevent Sybil attacks in a network, a variety of solutions have been put forward by various authors. However, there is still a need for improvements because the solutions currently offered have some drawbacks, such as the fact that some of the suggested solutions depend on numerous nodes or RSUs.\cite{b16}\cite{b13} have examined several security solutions in this article, and there are still several avenues that will be covered in this section for further investigation.
\subsection{Different types of cyber attacks}

RQ1: What are the different IoV risks, and whose security service are they affecting?

1. Sybil attack: In VANETs and IoV, the Sybil assault is one of the most common types of attacks. In conducting our research on the Sybil assault, we consulted recent reference publications. To prevent Sybil attacks in a network, a variety of solutions have been put forward by various authors. However, there is still a need for improvements because the solutions currently offered have some drawbacks, such as the fact that some of the suggested solutions depend on numerous nodes or RSUs.\\

2. Denial of service(DOS): Denial of Service (DoS) attacks are also considered to be among the most critical attacks and for our research, we used recent reference papers as sources. The following are some problems of the existing security solutions for DoS attacks: Few proposed approaches rely on manual parameter configuration, which is not a practical option, and in some proposed systems malicious nodes are not recognized if the network is overloaded with false information and genuine signatures. Therefore, even though there are many existing solutions, there is still room for development because of the previously stated limitations.\cite{b1}\\

3. Blackhole attack: A black-hole attack is a common attack in a vehicle system, and for our literature review, 18 reference publications from recent years were considered. Although there are several known security solutions for blackhole attacks, some of them have significant flaws, such as assuming the presence of a single malicious node, which is impossible in realistic conditions, or being operationally or memory intensive. To resolve the issue, new security measures must be developed to stop blackhole attacks.\\

4. Grayhole attack: The gray hole attack is one of the prominent attacks on in-vehicle networks and is difficult to detect due to the dual nature of the malicious node. About 10 research papers from the last few years were taken for our research, and existing security solutions have some drawbacks as many suggested methods only accept numerical numbers as input, and few depend on conventional routing protocols, which results in reduced efficiency, therefore new security solutions may be built by considering cryptographic techniques for guarding against grey hole attacks in a network.\\

5. Wormhole attack: The wormhole attack is one of the most notable attacks in IoV. For the wormhole attack, recent reference papers were considered to conduct our research and it has been determined that the existing security solutions to prevent these attacks have a few limitations such as excessive power consumption because of the large number of nodes, as well as results of proposed methods, are not compared with others. Because of these shortcomings, few other security solutions can be developed in the future to protect the networks from wormhole attacks.\\ 

6. Sinkhole attacks: Sinkhole attacks are also one of the most important attacks on in-vehicle networks and we have used 13 research papers from the last few years for this attack. Many security solutions have been created by various authors, but they all have some limitations, such as the inability to recognize malicious sinkhole nodes in the context of numerous attacks or when they are located close to base stations. Therefore, improvements to the current security solutions are required to protect the networks from such attacks.\\

7. Node impersonation attack: The node impersonation attack is also known, but the authors did not do much work to design the security solutions for this attack. recent reference papers were referenced during the survey and the alternatives that were presented are not very effective at securing the network and have a lot of overhead and delays. As a result, this attack needs to be considered, and effective security measures must be suggested.\\

8. Man-in-the-middle one: A popular attack in IoV is the man-in-the-middle one and 4 reference papers are used to conduct the research in this attack. Researchers gave this attack less consideration. In addition, a simulation for the suggested solution takes a lot of time. Hence, new security solutions should be set - up for this attack to protect the network.\\

\subsection{GPS (Global Positioning System) spoofing attacks}
GPS (Global Positioning System) spoofing attacks are also important attacks in the vehicular network. Three reference articles have been referred to for this attack. However, the writers haven't put much effort into developing security solutions. To stop this attack, it is necessary to develop more effective and safe solutions.\\

1. Masquerading attack: The masquerading attack is another challenging attack in IoV. There aren't many security solutions accessible for this assault, and only 5 reference articles were considered when researching security solutions. As a result, it is necessary to suggest cybersecurity for this attack to defend the network from masquerading attacks\\

2. Trust: Trust is an integral part of interacting entities, especially when they interact with strangers. For example, how sure can a node be that another node shares its data with them? Any trust system must have unique, persistent, and distinct identities for it to be viable. It is possible for nodes to change identities for subsequent interactions if they have non-persistent (with a short lifetime) identities, while non-distinct identities have no one-to-one mapping between identities and vehicles, that is, more than one identity on a single vehicle: Sybil attack. In the context of the law, the question might be: how can trust-related information be stored and managed on such a large scale? Or how can trust information be utilized securely? The future trust models developed should deliver identity assurance. Nodes cannot build trust on a network unless the trust system designers address this identity first. Interactions experiences also play an important role in establishing trust in the network. The models must be scalable and capable of distributed operation. Finally, they should ensure an accurate mapping of subjective to objective trust and be efficient in terms of overhead.\\

3. Resilience and Self-Adaptation: The transition from reducing vulnerabilities to enhancing resilience and self-adaptation is another significant direction that requires consideration. The IoV system should be strong enough to recover from attacks and unusual behaviors quickly and thoroughly. For robustness, researchers should investigate and use AI-based methods like automated software patching [3] and self-rewriting code [4] in the IoV domain.\\

4. Privacy preservation: The majority of IoV applications utilize cloud-based services. When it comes to delegated operations, third-party cloud-based services may not always be a trustworthy option. The majority of IoV applications leverage cloud-based services. Therefore, it is not always desirable to entrust delegated functions to outside cloud service providers. Utilizing cloud-based services without having your data disclosed would be more appealing. Currently used methods for protecting privacy in cloud data processing employ partially and completely advanced encryption standard algorithms. However, these algorithms consume a lot of resources, particularly when processing a sizable amount of data produced by several cars from the IoV environment. The IoV ecosystem requires lightweight fully homomorphic encryption to protect user and data privacy. Introduce controlled anonymity into the network as another option to consider protecting privacy. For example, on a cloud server, the users’ credentials must be required for authentication but should be anonymized passwords. Controlled anonymity means maintaining accountability and privacy while maintaining user anonymity. For example, users must not be so anonymous as to compromise accountability, but without being so anonymous as not to compromise privacy.\\

5. AI-based detection: There is a shortage of human resources for cybersecurity across the globe at present. The use of robotics and autonomous systems will increase in the future. It would be necessary to have an AI-based immune system that could deal automatically with unknown threats, protect against unseen threats and anomalies, and respond to AI-based malware, cognitive hackers, and so on. One such example is the IBM Watson project (https:// www.ibm.com/watson/).\\

In our research, we have discussed various attacks including Bogus Information Attacks, Replay Attacks, Spamming Attacks, Illusion Attacks, Snooping Attacks, ID Disclosure, Message Tampering, Brute Force Attacks, Social Attacks, Timing Attacks, Home Attacks, Repudiation Attacks, Session Hijacking, and Key and/or Certificate Replication Attack. Although these attacks are important, they have not received much attention from the research community and no security solutions have been proposed. So, it is necessary to design security solutions to protect the network from these attacks \cite{j3,j4}.\\

\subsection{Possible solution for different attacks}

RQ2: what is the possible solution for different security attacks in IoV.\\

1. Solution for Sybil attack: To detect malicious nodes in Sybil attacks, authors proposed an approach called obfuscated neighbor relationships between roadside units (DMON). The four objectives of the proposed method are: \\

(a). A ring signature-based identification scheme based on neighbor relationships is used to detect malicious Sybil nodes by generating a signed certificate as a temporal identity. \\

(b). RSU neighboring relationships are perplexed to protect vehicle privacy information.\\

(c). The ability to detect online and independently. \\

(d). Increasing efficiency while reducing overhead.\\

2. Solution for DOS attack: 12.	The authors propose an Enhanced Attacked Packet Detection Algorithm (PDA) as a security solution to DoS attacks. According to this scheme, each RSU is examined using the EAPDA algorithm, which compares the rate at which each RSU sends data packets. If RSUS notices that malicious nodes are transferring more packets than normal, those packets will be discarded.\\

3. Solution for Blackhole attack: The authors propose a method for verifying and detecting blackhole attacks using the backbone network. Nodes are classified into three categories based on the logic of this algorithm: This figure shows a backbone network with nodes outside the transmission range (LN), nodes strongly associated with the transmission range (HN), and nodes on the border of the transmission range (BN) \cite{b20}.\\

Additionally, there are two types of lists maintained here - an associated list with BN and LN, and an unassociated list without BN and LN. As a means of detecting malicious nodes, source nodes send information packet confirmation requests to the backbone network, and the backbone network, in turn, confirms packet delivery at the destination through end-to-end checks and destination nodes will not receive packets if malicious nodes are present. They will advise their clients to check their correspondence.\\

4. Solutions for Grayhole attack: Gray hole attacks can be minimized by employing the IMP (IMProvement) method, which focuses on improving Denial Contradictions with Fictitious Nodes (DCFM). The Multi-Point Relay Node (MPR) is defined as a node that performs routing decisions based on two contradiction rules introduced in DCFM, and other nodes can also apply the defined rules to take router decisions\\

5. Solution for SInk hole attack: To detect malicious nodes in the network, the authors have proposed a method based on the energy consumption model in the AODV routing protocol using pure MD5. A pure MD5 system requires both public and private keys to ensure security. Each node is granted a signature to enhance security. If energy is reduced by using an external battery, energy is supplied to that node. Detection of attacker signatures by the proposed algorithm and determination of the shortest route to sink is provided if the signature of the attacker node matches another legitimate node's signature.\\

6. Solution for Node Impersonation method: Using RSS inherited from remote hubs for assault identification, the authors propose a framework for detecting and preventing impersonation attacks. This approach does not depend on cryptography to detect mocking attacks, since it uses unsupervised threshold methods to split RSS hints into two classes.\\

7. Solution for MAN-in-the-middle one: A location-based pseudonym and a hash function used in combination with the SHA-1 algorithm prevented the man-in-the-middle attacks in Anonymous Location-Based Efficient Routing Protocol (ALERT). For packet delivery to be successful, the proposed scheme incorporates ACKS, so malicious nodes can be detected by computing hash values at both sources and destinations, and will, in the case of detection, send a negative acknowledgment to the sender packets again and use an alternative path for routing.\\

8. Solution for Masquerading attack: To protect RSS readings from being faked, the authors have implemented an anomaly detection model that uses signal strength fluctuation to detect masquerade attacks. As part of the detection mechanism, two concepts are considered for detecting malicious nodes. The first is the time of reception of each RSS and the second is the maximum speed of any node.\\

\section{Future research direction of the internet of vehicles security}

Modern intelligent transportation systems have benefited the world to a large extent. On the other side safety challenges also toped up for IoV-based Machines. Data integrity, reliability, and accuracy of AI-based models, and effective traffic management need to be ensured for a reliable, secure, and effective future of IoT. In this section, we will present modern aspects of the internet of vehicles with their challenging requirements.\\

\subsection{AI-based autonomous vehicles}

Intelligent transportation systems that can independently navigate the environment and improve the reliability of autonomous vehicles while making them safer on the road are accomplished by the integration of advanced sensing, AI models, and edge computing into the Internet of Vehicles system. Models receive data from affiliated users and predict specific decisions based on accuracy standards. The real-world AI-based IoV demands accurate and reliable predictions for safe and effective transportation. The emerging use of AI technologies which performs all task effectively and independently with encrypted data transmission brings the challenge of confirmed accuracy requirements that can prevent fatal road accidents and provide safer and more comfortable transportation.\cite{b16}\cite{b4,j4,j5}.

\subsection{Block chain enabled IoV.}
Blockchain is a series of blocks that each hold transactions, data, and scripts. Using various cryptographic methods, all the blocks are connected to form a chain. In a blockchain system that is frequently updated by all network participants, the newly created blocks are persistently linked to the chain. The main objective of blockchain technology is to provide security to IoV data for future analysis. Blockchain enables the transformation of IoV data to a centralized cloud-based server by enabling trusted and robust transactions in a common and insecure network. These advancements require efficient security of data transactions between nodes. This demands more effective advancements in terms of the security of the Internet of vehicles.
\subsection{Effective edge intelligence for the IoV system.}
An intelligent IoV system can provide comfortable, and safe transportation by providing multiple features at the same time. This includes video streaming, traffic management, auto-driving, various communications tools, and more. On the other side, this also features the present challenge of quick and correct responses in all scenarios. Data analysis and minimum delay in response are the key deliverables for edge intelligent networks. 
\subsection{Federated learning solutions for IoV}
A basic machine learning algorithm performs in a standard way that receives data and training through the model. The classified server gets data from IoT sensors and gives an accurate measurement. Third-party servers can become an active entities while data transferring which leads to data breaching and unreliable systems. Federated learning is an emerging system that sends only notable updates to the local ML model between common devices.\cite{b26} Federated learning enables a secure transportation medium for models as it performs only local training on data at the remote computing devices which is more secure and efficient compared to the basic ML approach.
The following are questions that should be answered for an effective and secure IoV system.\\
• How can we identify the best AI-based approach to ensure maximum prediction accuracy?\\
• How to secure data transfer between common nodes in blockchain-based IoV?\\
• How can we minimize the response delay in heavy traffic conditions and vehicular communication?\\
\section{Conclusion and Future work}
This research presents an overall view of the internet of vehicles and its security measures. The purpose of this research is to identify possible threats of IoV and its solution. The initial phase of this research represents the finding of related work and architecture of IoV. IoV has a vast array of applications that will ensure a bright future for IoV with future research advancements. Research also shows that there are many aspects of IoV that are vulnerable and how the overall security can be compromised as well as a possible solution to each threat. IoV is still in its infancy and has a large field to discover with its emerging intelligent advancement which will ensure demanded security for IoV.
This study shows the opportunity available for future research within the area of IOV which will ensure effective, safe, and reliable transportation. As numbers are heavily increasing for vehicles it poses a challenge for every developing technology within the area of the automobile.\\

Research Agenda 1: IoV has developed to a large extent but, there are future opportunities available in the context of efficiency of information routing and dissemination. VANETs present multiple MAC address-based communication, still, there is multi-level communication that takes place during transportation demands efficient communication protocols. Studies and research are in progress to advance this area of IoV.\\

Research Agenda 2: there are numerous advancements in terms of functionality, communication, and intelligent behaviors in IoV. Fuel saving and performance optimization remain undiscovered parts of IoV. AI is advancing in every aspect of modern technology, Fuel optimization, and environment-friendly emission is in the beginning stage of research.\\

Research Agenda 3: Traffic data management is a key concern in the safety and efficiency of IoV-enabled vehicles. Intersection control, traffic light analysis, and surrounding assessment are important aspects of Traffic Management. Research is dividing traffic management into highway traffic and city traffic. City traffic is complicated in terms of assessment compared to highway traffic. \\

\vspace{12pt}
\color{red}

\end{document}